\documentclass[a4paper]{article}

\usepackage{t1enc}
\usepackage{amsmath}
\usepackage{amssymb}

\newcommand{\edth} {\mbox{\symbol{'360}}}
\newcommand{\dimker}{{\dim\ker}}
\newcommand{\2}{\frac12}
\newcommand{\4}{\frac14}
\newcommand{\NN}{{\mathbb N}}
\newcommand{\RR}{{\mathbb R}}
\newcommand{\CC}{{\mathbb C}}
\newcommand{\ZZ}{{\mathbb Z}}
\renewcommand{\S}{{\mathcal S}}

\numberwithin{equation}{section}

\title{The kernel of the edth operators on higher-genus spacelike 
two-surfaces}
\author{J\"org Frauendiener\\
MPI f\"ur Mathematik in den Naturwissenschaften\\
D-04103 Leipzig\\
Inselstrasse 22--26\\
Germany
\and
L\'aszl\'o B. Szabados\\
Research Institute for Particle and Nuclear Physics\\
H-1525 Budapest 114\\
PO Box 49\\
Hungary
}

\begin{document}
\maketitle

\begin{abstract}
The dimension of the kernels of the edth and edth-prime operators 
on closed, orientable spacelike 2-surfaces with arbitrary genus is
calculated, and some of its mathematical and physical 
consequences are discussed. 
\end{abstract}
\newpage

\section{Introduction}

The ``edth'' operator $\edth$ made its first appearance in General
Relativity in an article by Newman and
Penrose~\cite{newman66:_note_bondi_metzn_sachs} on the symmetry group
of asymptotically flat spacetimes. It was introduced as a particular
differential operator on the unit-sphere acting on spin-weighted
functions, i.e., sections of certain complex line bundles over the
sphere. In this paper and~\cite{goldbergmacfarlane67:_spin} the
relationship of $\edth$ and its complex conjugate counterpart $\edth'$
with the representation theory of the Lorentz group resp. $SL(2,\CC)$
has been elaborated. In particular, the so called spin weighted
spherical harmonics on the sphere were introduced. Since then, the
edth operator has been intensely studied and it has found numerous
applications in General Relativity because it provides a useful tool
for all problems which lead to elliptic tensor or spinor equations on
the sphere~\cite{gerochheld73:_ghp}.

In~\cite{eastwoodtod82:_edth} it was pointed out that $\edth$ and
$\edth'$ are closely related to the $\partial$ and $\bar\partial$
operators of complex analysis. In fact, this analogy can be used to
define $\edth$ and $\edth'$ on arbitrary compact Riemann surfaces, the
sphere being merely the special case of a surface with genus
$g=0$. The natural framework for this definition of $\edth$ and
$\edth'$ is the theory of complex line bundles over Riemann surfaces.

Riemann surfaces with higher genus have many uses in conformal field
theories and string theory (see
e.g.~\cite{alvarez-gaumenelson86:_rieman,schlichenmaier89:_rieman} for
an introduction). In particular, the discussion of higher genus black
holes
(e.g.,~\cite{woolgar99:_bound,gallowayschleich99:_topol_censor_higher_genus_black_holes})
has recently gained importance due to the adS/CFT conjecture which
asserts that supergravity in an asymptotically adS (anti-de Sitter)
spacetime corresponds to a conformal field theory on the conformal
boundary of this spacetime in a certain limit. Such spacetimes admit
black holes with non-trivial topology which is accordingly reflected
in the topology of their conformal boundary.

But also in the classical theory of gravity do they make an
appearance. E.g., when the Einstein equations have been reduced by
symmetry assumptions to the Ernst equation, a completely integrable
system, then one can employ the methods known from soliton theory to
produce exact solutions of the field equations which are given in
terms of higher genus Riemann surfaces and their associated
theta-functions (see
e.g.~\cite{meinel00:_gravit,kleinrichter99:_exact,frauendienerklein:_ii_physic_proper}
for some applications).

Our immediate motivation for the present paper arose from two
different sources. Schmidt~\cite{schmidt96:_vacuum} has presented a
solution of the vacuum Einstein equations which has a conformal
boundary with toroidal cross sections. This solution has been very
useful for tests of numerical codes solving Friedrich's conformal
field
equations~\cite{huebner99:_scheme_einstein,frauendiener98:_numer_hivp_ii}.
The discussion of the asymptotic properties like e.g., their Bondi
4-momentum, radiation field and asymptotic symmetry group, of this and
more general such spacetimes necessarily leads to the examination of
$\edth$ equations on higher genus Riemann
surfaces~\cite{frauendiener99:_habil}.  One particular problem which
arises in this context is the question as to whether the Weyl tensor
also necessarily vanishes on the conformal boundary of these
spacetimes. Recall~\cite{penrose:_spinor_spacet}, that this
question can be answered to the affirmative in the spherical case
because a certain $\edth$ equation has no non-trivial solution.

Another motivation comes from the construction of two-surface
observables (e.g. quasi-local energy-momentum and angular momentum)
associated with spacelike 2-surfaces embedded in a
spacetime~\cite{szabados:_quasi_i}. While the quasi-local constructions 
have been mainly carried out for the spherical case, there is no reason 
to believe that this should be the only possibility. Again, one is lead 
to solve equations of the type $\edth\phi=\omega$ for given $\omega$ on 
a closed spacelike 2-surface $\S$. However, the structure of the
solutions depends on the kernel and cokernel of the operator $\edth$,
which in turn depends on the topology of the line bundle $E$ whose cross 
section is $\phi$. Since the topology of $E$ depends on the topology 
of $\S$, to solve certain physical problems one should know the kernel 
and cokernel of the edth operators on complex line bundles $E$ over 
2-surfaces that have more complicated topology than that of $S^2$.

In this paper, we do not intend to discuss the general theory of such
operators on complex line bundles. These issues are discussed in the
basic monograph~\cite{gunning66:_lectur_rieman_surfac} in
general. Rather, we concentrate on the applicability of our results to
General Relativity, motivated by the problems above. Thus, the line
bundles that we consider are connected with the spacetime structure.
They arise as the bundles associated with the specific bundle of the
spacetime spin frames over the 2-surface $\S$. We regard them and
various structures on them as being derived from the four-dimensional
geometry.

The necessary background on the geometry of spacelike 2-surfaces and,
in particular, the spinor geometry of $\S$ can be found in
\cite{szabados94:_two_sen_relat,szabados94:_two_sen_quasi_loc,szabados97:_quasi_sen},
which also contain the calculation of $\dimker\,\edth_{(p,q)}$ and
$\dimker\,\edth'_{(p,q)}$ for $\S\approx S^2$. Essentially that
technique is used here to calculate these kernels also for higher-genus
2-surfaces.

Our conventions are those of~\cite{penrose:_spinor_spacet} while our use 
of global differential geometry is based 
on~\cite{warner83:_found_differ_manif_lie_group,besse87:_einst_manif,kobayashinomizu:_found_differ_geomet}.

\section{The geometry of closed spacelike 2-surfaces}

It is known that for any connected, closed, orientable two dimensional
smooth manifold $\S$ the de Rham cohomology spaces are
$H^0(\S)=H^2(\S)=\RR$ and $H^1(\S)=\RR^g$, where $g$ is the genus of
$\S$. Let $\{a_i,b_i\}$, $i=1,\ldots ,g$, be a canonical homology
basis on $\S$. These are closed non-contractible curves on $\S$, and 
for the fundamental group of $\S$ we have
\[
\pi_1(\S)=\langle a_i,b_i\vert \,i=1,\ldots ,g;\,\, \prod^g_{i=1}
a_ib_ia^{-1}_ib^{-1}_i=1\,\rangle,
\]
i.e. $\pi_1(\S)$ is generated by the $2g$ elements $a_i$, $b_i$ with
the only relation that the product of all the commutants
$a_ib_ia^{-1}_ib^{-1}_i$ is homotopically
trivial~\cite{farkaskra93:_rieman_surfac}. Let $q_{ab}$ be a (negative
definite) metric on $\S$, $\varepsilon_{ab}$ the corresponding volume
2-form and $\delta_e$ the Levi--Civita covariant derivative. By the
Hodge decomposition every 1-form $\alpha_a$ is the sum of a closed, a
harmonic and an exact 1-form: $\alpha_a=\beta_a+\omega_a+\delta_af$,
where $\delta_{[a}\beta _{b]}=0$, $\delta_{[a}\omega_{b]}=0$ and
$\delta_a\omega^a=0$ hold, and $f:\S\rightarrow\RR$ is some
function. Furthermore, this decomposition is unique, and hence, there
are $2g$ linearly independent harmonic 1-forms on
$\S$~\cite{warner83:_found_differ_manif_lie_group}. Any basis
$\{\alpha^i_a,\beta^i_a\}$, $i=1,...,g$, in the space of the (real)
harmonic 1-forms can be uniquely characterized by the four $g\times g$
matrices $M^i{}_j:=\oint_{a_j}\alpha^i_a$,
$M^{g+i}{}_j:=\oint_{a_j}\beta^i_a$,
$M^i{}_{g+j}:=\oint_{b_j}\alpha^i_a$ and
$M^{g+i}{}_{g+j}:=\oint_{b_j}\beta^i_a$, and the $2g\times2g$ matrix
$M^I{}_J$, $I,J=1,\ldots ,2g$, is nonsingular\footnote{The $2g$ real
harmonic forms can be combined to form $g$ holomorphic 1-forms, the so
called Abelian differentials of the first kind. The real matrix $M$
then corresponds to the complex $g \times 2g$-matrix of periods of the
holomorphic differentials. This matrix can always be put into the form
$\left(1_g \vert P\right)$, where $P$ is a symmetric complex matrix
with positive definite imaginary part, the so called Riemann
matrix~\cite{farkaskra93:_rieman_surfac}.}. Another set of 1-forms,
associated naturally with $(\S,q_{ab})$ are the conformal Killing
1-forms. They are defined by the vanishing of the symmetric trace-free
part of their covariant derivative:
$\delta_{(a}\xi_{b)}-{\frac12}q_{ab}\delta_e \xi^e=0$. \par

Next, suppose that $\S$ is embedded as a smooth spacelike submanifold
in $M$ (for a detailed discussion of what follows, see
\cite{szabados94:_two_sen_relat,szabados94:_two_sen_quasi_loc,szabados97:_quasi_sen}). If $t^a$ and $v^a$ are future timelike and spacelike
unit normals to $\S$ satisfying $t^av_a=0$, respectively, then
$\Pi^a_b:=\delta^a_b-t^at_b+v^av_b$ is the orthogonal projection to
$\S$, and $O^a_b:=\delta^a_b-\Pi^a_b$ is the projection to the
timelike 2-planes orthogonal to $\S$. Such vector fields are globally
well defined if $\S$ is orientable and at least an open neighbourhood
of $\S$ in $M$ is time and space orientable. The induced metric on
$\S$ is $q_{ab}:=\Pi^c_a\Pi^d_bg_{cd}$, and if $\varepsilon_{abcd}$ is
the natural volume form on $M$, then the induced area 2-form and area
element on $\S$ is $\varepsilon_{cd}:=t^av^b \varepsilon_{abcd}$ and
${\rm d}\S:=\2\varepsilon_{cd}$, respectively. The area 2-form on the
orthogonal 2-planes is ${}^\bot \varepsilon_{ab}:=t_av_b-t_bv_a$. Then
any four-vector $X^a$ at the points of $\S$ can be decomposed in a
unique way into the sum of its tangential and normal parts as
$X^a=\Pi^a_bX^b+O^a_bX^b$. This implies that ${\bf V}^a(\S)$, the
restriction to $\S$ of the tangent bundle $TM$ of $M$, has the
$g_{ab}$-orthogonal decomposition ${\bf V}^a(\S)=T\S\oplus N\S$ into
the sum of the tangent bundle $T\S$ and the (globally trivializable)
normal bundle $N\S$ of $\S$. By the orientability of $\S$ the bundle
of $q_{ab}$-orthonormal frames in $T\S$ is reducible to a
$B(\S,SO(2))$ principal bundle, while the time and space orientability
of $(M,g_{ab})$ implies that the bundle of orthonormal frames in the
normal bundle $N\S$ is reducible to a $B(\S,SO(1,1))$ principal
bundle. While the latter is always trivial, the former is not. In
particular, if $g=0$, i.e. $\S\approx S^2$, then $B(\S,SO(2))\approx
\RR P^3$, but for $g=1$, i.e. for $\S\approx S^1\times S^1$,
$B(\S,SO(2))\approx\S\times SO(2)\approx S^1\times S^1 \times
S^1$. Therefore, the bundle of $g_{ab}$-orthonormal frames with given
time and space orientation that are compatible to the decomposition ${\bf
V}^a(\S)=T\S\oplus N\S$ is $B(\S,SO(2) \times SO(1,1))\approx
B(\S,SO(2))+B(\S,SO(1,1))$.

The spacetime Levi-Civita covariant derivative $\nabla_a$ defines a
covariant derivative on ${\bf V}^a(\S)$ by $\delta_eX
^a:=\Pi^a_b\Pi^f_e\nabla_f\bigl(\Pi^b_cX^c\bigr)+O^a_b\Pi^f_e\nabla_f
\bigl(O^b_cX^c\bigr)$. This derivative annihilates both the spacetime
metric and the projections $\Pi^a_b$ and $O^a_b$; and hence
annihilates the intrinsic metric $q_{ab}$. Furthermore, it is
symmetric: $(\delta_c\delta_d-\delta_d\delta_c)\phi=0$ for any smooth
function $\phi$ on $\S$. Geometrically, $\delta_e$ is the covariant
derivative on ${\bf V}^a(\S)$ determined by a connection on the sum of
the principal $SO(2)$-- and $SO(1,1)$ bundles above. The commutativity
of the structure group $SO(2)\times SO(1,1)$ implies that $\delta_e$
does not `mix' the tangential and normal sections. In particular, the
connection coefficient corresponding to the vertical part of the
connection can be represented by the 1-form $A_a:=\Pi^b_a(\nabla_bt_c)
v^c=v^c\delta_at_c$. The curvature of $\delta_a$, defined by
$-f^a{}_{bcd}X^b:=(\delta_c\delta_d-\delta_d\delta_c)X^a$, is
\begin{equation}
f^a{}_{bcd}=-{}^\bot\varepsilon^a{}_b\Bigl(\delta_cA_d-\delta_dA_c
\Bigr)+\2\,{}^\S R\Bigl(\Pi^a_cq_{bd}-\Pi^a_dq_{bc}\Bigr),
\label{(2.1)}
\end{equation}
where ${}^\S R$ is the curvature scalar of the intrinsic curvature 
of $(\S,q_{ab})$. This is the sum of the curvatures corresponding 
to the $SO(1,1)$-connection on $N\S$ and the $SO(2)$-connection 
on $T\S$. \par

Suppose that $(M,g_{ab})$ admits a spinor structure, and let ${\bf
S}^A(\S)$ be the pull back to $\S$ of the bundle of unprimed spinors
over $M$. Let $t_{AA'}$, $v_{AA'}$ be the spinor form of the normals
to $\S$, and let us define $\gamma^A{}_B:=2t^{AA'} v_{BA'}$. This
defines the bundle automorphism $\gamma:{\bf S}^A(\S)\rightarrow{\bf
S}^A(\S):\lambda^A\mapsto\gamma^A{}_B\lambda ^B$, and
$\gamma^A{}_A=0$, $\gamma^A{}_B\gamma^B{}_C=\delta^A_C$ hold. Thus
$\gamma^A{}_B$ has two eigenvalues, $\pm1$, and $\pi^{\pm A}{}_B
:=\2(\delta^A_B\pm\gamma^A{}_B)$ are the projections of ${\bf S}
^A(\S)$ to the bundle ${\bf S}^{A\pm}(\S)$ of the $\pm1$ (right
handed/left handed) eigenspinors, respectively. If $o^A$ is a left
handed and $\iota^A$ is a right handed spinor normalized by $o_A
\iota^A=1$, then they form a GHP spinor dyad~\cite{gerochheld73:_ghp}
on $\S$. The projection $\Pi^a_b$ can be expressed by $\gamma^A{}_B$,
too: $\Pi^a_b=\2(\delta^A_B\delta^{A'}_{B'}-\gamma^A{}_B\bar\gamma^{A'}
{}_{B'})$. Thus the area 2-form on $\S$ and on the orthogonal 2-planes,
respectively, are $\varepsilon_{cd}=\frac{\rm i}{2}(\gamma
_{CD}\varepsilon_{C'D'}-\bar\gamma_{C'D'}\varepsilon_{CD})$ and ${}
^\bot\varepsilon_{cd}=-\2(\gamma_{CD}\varepsilon_{C'D'}+\bar
\gamma_{C'D'}\varepsilon_{CD})$. $\gamma_{AB}=\gamma_{(AB)}=2o_{(A}
\iota_{B)}$ can also be considered as a complex fibre metric on ${\bf
S}^A(\S)$. $\pi^{\pm a}{}_b:=\pi^{\pm A}{}_B\bar\pi^{\mp A'}{} _{B'}$
are the projections of the complexified tangent bundle $T\S
\otimes\CC$ (in fact, the complexified Lorentzian vector bundle ${\bf
V}^a(\S)\otimes\CC$) to the sub-bundles $T^{(1,0)}\S$ and
$T^{(0,1)}\S$ of the complex tangent vectors spanned by
$m^a:=o^A\bar\iota^{A'}$ and $\bar m^a:=\iota^A\bar o^{A'}$,
respectively, i.e. $\gamma_{AB}$ determines the complex structure
\cite{kobayashinomizu:_found_differ_geomet} of $\S$, too. The
analogous projections $o^{\pm a}{}_b:=\pi^{\pm A}{}_B\bar\pi^{\pm
A'}{}_{B'}$ define respectively the two null normals
$n^a:=\iota^A\bar\iota^{A'}$ and $l^a:=o^A\bar o^{A'}$ up to
scale. The bundle of the GHP spinor dyads is a principal bundle
$\tilde B(\S,GL(1,\CC))$, which is a double covering bundle of $B(\S,
SO(2)\times SO(1,1))$. In particular, if $\S$ is a topological
2-sphere, then $\tilde B(\S,GL(1,{\bf C}))\approx S^3\times (0,
\infty)$, and since $B(\S,SO(2)\times SO(1,1))$ is trivial for $g=1$,
its covering $\tilde B(\S,GL(1,{\bf C}))$ is also trivial. The
derivative $\delta_e$ extends naturally to ${\bf S}^A(\S)$, by the
requirement that it should annihilate both $\varepsilon_{AB}$ and 
$\gamma_{AB}$. For its curvature we obtain
\begin{equation}
f^A{}_{Bcd}=-\2\gamma^A{}_B\Bigl(\bigl(\delta_cA_d-\delta_dA_c
  \bigr)-\4\,{}^\S R\bigl(\varepsilon_{C'D'}\gamma_{CD}-
  \varepsilon_{CD}\bar\gamma_{C'D'}\bigr)\Bigr). \label{(2.2)}
\end{equation}
We can define the curvature scalar of $\delta_e$ by $f:=f_{abcd} \2
(\varepsilon^{ab}-{\rm i}\,{}^\bot\varepsilon^{ab})\varepsilon
^{cd}={\rm i}\gamma^A{}_Bf^B{}_{Acd}\varepsilon^{cd}={}^\S R-2{\rm
  i}\delta_c(\varepsilon^{cd}A_d)$, which is four times the 
so-called {\it complex} Gauss curvature of~\cite{penrose:_spinor_spacet}. 
Then by the Gauss--Bonnet theorem $\oint_\S f\,{\rm d}\S=\oint_\S{}^\S 
R\,{\rm d}\S=8\pi(1-g)$, and hence $\delta_e$ can be flat only if $\S$ 
is a torus.

\section{The line bundles $E(p,q)$}

Let $\rho_{(p,q)}:\,GL(1,\CC)\times\CC\rightarrow\CC:$
$(\lambda,z)\mapsto\lambda^{-p}\bar\lambda^{-q}\,z$, which is a left
action of $GL(1,\CC)$ on $\CC$ precisely when $p-q\in\ZZ$
(although $p+q$ may be arbitrary complex number, we assume that it is
real), and let $E(p,q)$ denote the complex line bundle associated to
$\tilde B(\S,GL(1,\CC))$ with the group representation $\rho
_{(p,q)}$. $E(p,q)$ is called the bundle of $(p,q)$-type scalars over
$\S$. They have the following elementary properties
\cite{eastwoodtod82:_edth,porter83:_edth,baston84:_index_twist} :
\begin{enumerate}
\item The complex conjugate bundle is $\overline{E(p,q)}=E(q,p)$.
\item The tensor product bundle is $E(p,q)\otimes E(r,s)=E(p+r,q+s)$.
\item $E(p,p)$ is a trivial vector bundle for any $p\in\RR$, and all
  the $E(p,q)$'s are trivial over the torus $\S\approx S^1\times S^1$.
  Note that the trivial vector bundles admit nowhere vanishing sections. 
\item For any fixed, nowhere vanishing section $h$ of $E(1,1)$,
  \begin{equation}
  \langle\phi,\psi \rangle_{(p,q)}:=\oint_\S\vert
  h\vert^{-(p+q)}\phi\bar\psi\,{\rm d} \S
\label{(3.1)}
\end{equation}
  defines a positive definite
  Hermitian scalar product on the (infinite dimensional) vector space
  $E^\infty(p,q)$ of the smooth sections of $E(p,q)$. Such a section
  might be, for example, the everywhere positive function
  $h:=t_{AA'}o^A\bar o^{A'}$.
\item $E^\infty(0,0)=C^\infty(\S,\CC)$, and for the spinor bundle
  ${\bf S} ^A(\S)\rightarrow E(1,0)\oplus E(-1,0):$
  $\omega^A\mapsto(\omega ^Ao_A,\omega^A\iota_A)$ is a vector bundle
  isomorphism.
\item For the complexified tangent and normal bundles $T\S\otimes\CC
  \rightarrow E(1,-1)\oplus E(-1,1):$ $X^a\mapsto(X^am_a,X^a\bar m_a)$
  and $N\S\otimes\CC\rightarrow E(1,1)\oplus E(-1,-1):$ $V^a
  \mapsto(V^al_a,V^an_a)$ are vector bundle isomorphisms.
\item For the complex structure $\Lambda^{(1,0)}T^*\S\simeq
  T^{(0,1)}\S \rightarrow E(1,-1):$ $\tilde z^a\mapsto\tilde z^am_a$
  and $\Lambda^{(0,1)}T^*\S\simeq T^{(1,0)}\S\rightarrow E(-1,1):$
  $z^a \mapsto z^a\bar m_a$ and are vector bundle isomorphisms.
  $\Lambda^{(1, 0)}T^*\S$, spanned by the differential of the {\it
    holomorphic} coordinates, is called the canonical bundle of the
  Riemann surface $\S$ \cite{gunning66:_lectur_rieman_surfac}.
\end{enumerate}
The covariant derivative $\delta_a$ on ${\bf S}^A(\S)$ 
defines a covariant derivative $\edth_a$ of $\phi\in E^\infty(p,q)$ for 
any $p,q\in\ZZ$ by 
\begin{equation}
\edth_e\phi:=\delta_e(\phi\iota^{A_1}\ldots \iota^{A_m}o_{B_1}\ldots o
_{B_n}\bar\iota^{C'_1}\ldots \bar\iota^{C'_r}\bar o_{D'_1}\ldots \bar 
o_{D'_s})o_{A_1}\ldots o_{A_m}\iota^{B_1}\ldots \iota^{B_n}\bar o_{C'
_1}\ldots \bar o_{C'_r}\bar\iota^{D'_1}\ldots \bar\iota^{D'_s},
\label{(3.2)}
\end{equation}
where $m,n,r,s=0,1,2,\ldots $ such that $p=m-n$ and $q=r-s$. If $p\in
(0,1)$, then let $\edth_a$ be the covariant derivative on $E^\infty
(p,p)$ for which $\edth_a(\phi^{1/p})=\frac1p\phi^{\frac{1-p}{p}}\edth
_a\phi$ for any $\phi\in E^\infty(p,p)$. Then by the Leibniz rule and 
property 2. above $\edth_a$ can be extended to $E(p,q)$ for any $p,q
\in\RR$ satisfying $p-q\in \ZZ$. 
The curvature of $\edth_a$ on $E(p, q)$, defined by $-{\cal F}_{ab}v^a
w^b\phi:=v^a\edth_a(w^b\edth_b\phi)-w^a\edth_a(v^b\edth_b\phi)-[v,w]^a
\edth_a\phi$, is ${\cal F}_{ab}=\frac{\rm i}{4} (-pf+q\bar f)
\varepsilon_{ab}$, and hence the integral of the first Chern class of 
$E(p,q)$ is $c_1(p,q):=-\oint_\S\frac{\rm i}{4\pi}{\cal F}_{ab}=-(p-q)
(1-g)$. In particular, the vanishing of $c_1(p, q)$ characterizes the 
trivial line bundles, and for the canonical bundle it is $c_1(1,-1)=2
(g-1)$, in accordance with \cite[pp.110]{gunning66:_lectur_rieman_surfac}. 

The covariant directional derivatives, $\edth\phi:=m^e\edth_e\phi$ and
$\edth'\phi:=\bar m^e\edth_e\phi$, the edth and edth-prime operators, 
are elliptic differential operators
$\edth:E^\infty(p,q)\rightarrow E^\infty(p+1,q-1)$ and $\edth':E^\infty 
(p,q)\rightarrow E^\infty(p-1,q+1)$, and hence, because of the 
compactness of $\S$, their kernel is finite dimensional
\cite{warner83:_found_differ_manif_lie_group,besse87:_einst_manif}. 
The formal adjoint of the edth and edth-prime operators with respect
to the Hermitian scalar product $\langle .\vert. \rangle_{(p,q)}$
above are 
\begin{equation}
\begin{split}
  ({\edth}_{(p,q)})^\dagger&=-\vert h\vert^{(p+q)}
  {\edth}'_{(-q+1,-p-1)}
  \vert h \vert^{-(p+q)}, \\
  ({\edth}'_{(p,q)})^\dagger&=-\vert h\vert
  ^{(p+q)}{\edth}_{(-q-1,-p+1)}\vert h\vert^{-(p+q)}.
\end{split}
\label{(3.3)}
\end{equation}
They are also elliptic, and hence the analytic index of the edth and
edth-prime operators, defined by ${\rm ind}(\edth_{(p,q)}):=\dimker\,
\edth_{(p,q)}-\dimker\,(\edth_{(p,q)})^\dagger$ and ${\rm ind}(\edth'
_{(p,q)}):=\dimker\,\edth'_{(p,q)}-\dim\ker\,(\edth'_{(p,q)})^\dagger$, 
are finite \cite{warner83:_found_differ_manif_lie_group,
besse87:_einst_manif}. In terms of these notions the Riemann--Roch 
theorem of the theory of Riemann surfaces 
\cite[pp.111]{gunning66:_lectur_rieman_surfac} takes the form 
\begin{equation}
\begin{split}
  {\rm ind}(\edth_{(p,q)})&=(1+p-q)(1-g), \\
  {\rm ind}(\edth'_{(p,q)})&=(1-p+q)(1-g),
\end{split}
\label{(3.4)}
\end{equation}
which are just Baston's formulae~\cite{baston84:_index_twist}.

For any real 1-form $\omega_a$ one has $\omega:=\omega_am^a \in
E^\infty(1,-1)$ and $\bar\omega:=\omega_a\bar m^a\in E^\infty(-1,1)$;
and $2\delta_{[a}\omega_{b]}m^a\bar m^b=\edth\bar\omega-\edth'\omega$,
$2\delta_{(a}\omega_{b)}m^a\bar m^b=\edth\bar\omega+\edth'\omega$ and
$(\delta_a\omega_b)m^am^b=\edth\omega$ hold. Therefore, $\omega_a$ is a
real harmonic 1-form iff $\edth'\omega=0$, and $\omega_a$ is (the dual
of) a real conformal Killing vector iff $\edth\omega=0$.

A cross section $\phi$ of $E(p,q)$ is called
holomorphic/anti-holomorphic if $\edth'\phi=0$ or $\edth\phi=0$,
respectively. The point $P\in \S$ is said to be the zero of the
holomorphic section $\phi$ with order $m$ if $\phi$ and its first
$(m-1)$ $\edth$-derivatives vanish at $P$, but its $m$th
$\edth$-derivative is not zero there. It is not difficult to show that
$P$ is a zero of the holomorphic cross section $\phi$ with order $m$
if and only if there is a holomorphic section $\psi$ and a holomorphic
function $f$ on some open neighbourhood $W\subset\S$ of $P$ such that
$\phi=f\psi$ and $\psi$ is nonzero on $W$ and $f$ has a single zero
with order $m$ at $P$. The meromorphic/anti-meromorphic sections are
defined analogously with the only difference that we allow them to
have isolated poles. The order of the pole $P\in\S$ of the meromorphic
section $\phi$ is defined to be $n$ if for some holomorphic function
$f$, defined on an open neighbourhood $W$ of $P$ and having $P$ as a
zero with order $n$, $f\phi$ is holomorphic on $W$, while it is not 
holomorphic for functions $f'$ whose zero at $P$ is only of order $n'<n$. 
Note that by the
compactness of $\S$ the meromorphic sections have finitely many zeros
and poles. The degree of a meromorphic section $\phi$ of $E(p,q)$ with 
zeros of order $m_1,\ldots ,m_k$ and poles of order $n_1,\ldots , n_l$ 
is defined by ${\rm deg}(\phi):=\sum^k_{i=1}m_i-\sum^l_{j=1}n_j$. By a 
theorem of the theory of Riemann surfaces 
\cite[pp.103]{gunning66:_lectur_rieman_surfac} this degree depends only
on the line bundle: ${\rm deg}(\phi)=c_1(p,q)$. In the analogous
statement on the anti-meromorphic sections $\psi$ of $E(p,q)$ one has 
the first Chern class of the complex conjugate bundle: ${\rm deg}(\psi)
=c_1(q,p)$.

\section{The calculation of $\dimker\,\edth_{(p,q)}$ and 
$\dimker\,\edth'_{(p,q)}$}

We calculate the dimension of the kernels using basically the technique 
of the appendix of \cite{szabados94:_two_sen_quasi_loc}. Thus first we 
recall the basic formulae. As a consequence of the similarity 
transformations \eqref{(3.3)} between $(\edth_{(p,q)})^\dagger$ and 
$\edth'_{(-q+1,-p-1)}$ and between $(\edth'_{(p,q)})^\dagger$ and 
$\edth_{(-q-1,-p+1)}$, the dimension of the kernel spaces of $(\edth
_{(p,q)})^\dagger$ and $\edth'_{(-q+1,-p-1)}$, and of $(\edth'_{(p,q)})
^\dagger$ and $\edth_{(-q-1,-p+1)}$ are the same. By substituting these 
into Baston's formulae \eqref{(3.4)} above, we obtain 
\begin{equation}
  \begin{split}
    \dimker\,\edth_{(p,q)}=&\bigl(1+p-q\bigr)(1-g) +
    \dimker\,\edth'_{(-q+1,-p-1)}, \\
    \dimker\,\edth'_{(p,q)}=&\bigl(1-p+q\bigr)(1-g) +
    \dimker\,\edth_{(-q-1,-p+1)}.
\end{split}
\label{(4.1)}
\end{equation}
By the Leibniz rule $\edth_{(p+p',q+q')}(\phi\psi)=\psi\edth_{(p,q)}
\phi+\phi\edth_{(p',q')}\psi$ for any $\phi\in E^\infty(p,q)$ and $\psi
\in E^\infty(p',q')$, one has the inequality 
\begin{equation}
\dimker\,\edth_{(p+p',q+q')}\ge \max\bigl\{\dimker\,\edth_{(p,q)}, 
\dimker\,\edth_{(p',q')} \bigr\}
\label{(4.2)}
\end{equation}
whenever $\bigl(\dimker \edth_{(p,q)}\bigr) \cdot \bigl( \dimker
\edth_{(p',q')} \bigr) \ne 0$. There is a similar inequality for the
edth-prime operator. Finally, the last ingredient that we need is
\begin{equation}
\dimker\,\edth_{(0,0)}=\dimker\,\edth'
_{(0,0)}=1,  \label{(4.3)}
\end{equation}
which is just the generalized Liouville theorem 
\cite[pp.6]{gunning66:_lectur_rieman_surfac}: Every 
holomorphic/anti-holomorphic function on a compact Riemann surface is 
constant. Then by \eqref{(4.2)} and \eqref{(4.3)} 
\begin{equation}
  \begin{split}
    &\bigl(\dimker\,\edth_{(p,q)}\bigr)\bigl(\dimker\,\edth_{(-p,-q)}
    \bigr)=0   \quad{\rm or} \quad \dim
    \,\ker\,\edth_{(p,q)}=\dimker\,\edth_{(-p,-q)}=1,\\
    &\bigl(\dimker\,\edth'_{(p,q)}\bigr)\bigl(\dimker\,\edth'_{(-p,-q)}
    \bigr)=0   \,\,\,\,\,{\rm or} \,\,\,\,\,\dim
    \ker\,\edth'_{(p,q)}=\dimker\,\edth'_{(-p,-q)}=1,
  \end{split}
\label{(4.4)}
\end{equation}
for any $p,q\in\RR$, $p-q\in\ZZ$. \par

\begin{figure}[htbp]
\begin{center}
  \setlength{\unitlength}{1mm}
\begin{picture}(120,50)(-60,-10)
\thicklines
\put(0,0){\line(0,1){30}}
\put(23,0){\line(0,1){30}}
\put(46,0){\line(0,1){30}}
\put(-23,0){\line(0,1){30}}
\put(-46,0){\line(0,1){30}}
\thinlines

\put(-69,-10){\makebox(0,0)[b]{$\cdots$}}
\put(-46,-10){\makebox(0,0)[b]{$E(p-2,q+2)$}}
\put(-23,-10){\makebox(0,0)[b]{$E(p-1,q+1)$}}
\put(  0,-10){\makebox(0,0)[b]{$E(p,q)$}}
\put( 23,-10){\makebox(0,0)[b]{$E(p+1,q-1)$}}
\put( 46,-10){\makebox(0,0)[b]{$E(p+2,q-2)$}}
\put( 69,-10){\makebox(0,0)[b]{$\cdots$}}

\put(-57.5,0){\vector(1,0){23}}
\put(-34.5,0){\vector(1,0){23}}
\put(-11.5,0){\vector(1,0){23}}
\put( 11.5,0){\vector(1,0){23}}
\put( 34.5,0){\vector(1,0){23}}
\put( 57.5,0){\line(1,0){11.5}}

\put(-34.5,-4){\makebox(0,0)[b]{$\edth$}}
\put(-11.5,-4){\makebox(0,0)[b]{$\edth$}}
\put( 11.5,-4){\makebox(0,0)[b]{$\edth$}}
\put( 34.5,-4){\makebox(0,0)[b]{$\edth$}}
\put( 57.5,-4){\makebox(0,0)[b]{$\edth$}}

\put( 57.5,30){\vector(-1,0){23}}
\put( 34.5,30){\vector(-1,0){23}}
\put( 11.5,30){\vector(-1,0){23}}
\put(-11.5,30){\vector(-1,0){23}}
\put(-34.5,30){\vector(-1,0){23}}
\put(-57.5,30){\line(-1,0){11.5}}

\put(-57.5,31){\makebox(0,0)[b]{$\edth^\dagger$}}
\put(-34.5,31){\makebox(0,0)[b]{$\edth^\dagger$}}
\put(-11.5,31){\makebox(0,0)[b]{$\edth^\dagger$}}
\put(11.5,31){\makebox(0,0)[b]{$\edth^\dagger$}}
\put(34.5,31){\makebox(0,0)[b]{$\edth^\dagger$}}

\linethickness{1pt}

\put(-46,5){\circle*{1}}
\put(-46,5){\line(4,-1){20}}
\put(-46,5){\line(0,-1){5}}
\put(-47,3){\makebox(0,0)[r]{1}}

\put(-23,5){\circle*{1}}
\put(-23,5){\line(4,-1){20}}
\put(-23,5){\line(0,-1){5}}
\put(-24,3){\makebox(0,0)[r]{1}}

\put(0,5){\circle*{1}}
\put(0,5){\line(4,-1){20}}
\put(0,5){\line(0,-1){5}}
\put(-1,3){\makebox(0,0)[r]{1}}

\put(23,5){\circle*{1}}
\put(23,5){\line(4,-1){20}}
\put(23,5){\line(0,-1){5}}
\put(22,3){\makebox(0,0)[r]{1}}

\put(46,5){\circle*{1}}
\put(46,5){\line(4,-1){20}}
\put(46,5){\line(0,-1){5}}
\put(45,3){\makebox(0,0)[r]{1}}

\put(-46,25){\circle*{1}}
\put(-46,25){\line(-4,1){20}}
\put(-46,25){\line(0,1){5}}
\put(-45,27){\makebox(0,0)[l]{1}}

\put(-23,25){\circle*{1}}
\put(-23,25){\line(-4,1){20}}
\put(-23,25){\line(0,1){5}}
\put(-22,27){\makebox(0,0)[l]{1}}

\put(  0,25){\circle*{1}}
\put(  0,25){\line(-4,1){20}}
\put(  0,25){\line(0,1){5}}
\put(  1,27){\makebox(0,0)[l]{1}}

\put( 23,25){\circle*{1}}
\put( 23,25){\line(-4,1){20}}
\put( 23,25){\line(0,1){5}}
\put( 24,27){\makebox(0,0)[l]{1}}

\put( 46,25){\circle*{1}}
\put( 46,25){\line(-4,1){20}}
\put(46,25){\line(0,1){5}}
\put( 47,27){\makebox(0,0)[l]{1}}

\end{picture}
\end{center}
\caption{\label{fig:torus}The sequence of line bundles and the edth and 
  the adjoint-edth operators on the torus. The vertical lines represent 
  the spaces of the smooth sections of the line bundles, and a piece of 
  them, the kernels, are mapped into the zero of the next space, but not 
  every section has a pre-image in the previous space. In particular, on 
  the torus the kernel and cokernel spaces of the edth operators are 1 
  dimensional. The kernel and cokernel of the adjoint-edth operator is 
  just the cokernel and kernel of the edth, respectively. }

\end{figure}
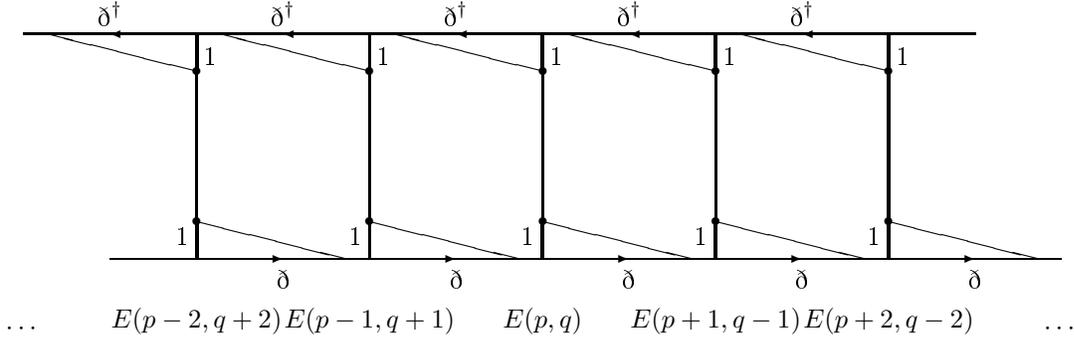

First let us consider the trivial bundles, i.e. $q=p$ or $g=1$, and
fix a holonomically trivial covariant derivative ${}_0\edth_a$ on
$E(p,q)$. [To see that such a connection exists, recall that there are
flat connections on the corresponding principal bundles, and their
holonomy groups, being homomorphic images of the fundamental group 
$\pi_1(\S)$ in the structure group, are discrete
\cite{kobayashinomizu:_found_differ_geomet}. Let $\tilde B$ be any
of these principal bundles, and $\tilde\omega_a$ the connection 1-form 
of the flat connection. Let us fix a canonical homology basis $\{a_i,
b_i\}$, $i=1, \ldots ,g$, of $\S$, let $\tilde h(a_i)$, $\tilde h(b_i)$ 
denote the corresponding holonomies, and define the connection 1-form
$\omega_a:= \tilde\omega_a+(A_i+{\rm i}B_i)\alpha^i_a+(C_i+{\rm
i}D_i)\beta^i_a$, where $A_i$, $B_i$, $C_i$ and $D_i$, $i=1,\ldots
,g$, are real constants, and let $\{\alpha^i_a,\beta^i_a\}$ be a basis
of the real harmonic 1-forms on $\S$. Then $\omega_a$ is also flat,
and the corresponding holonomies along $a_j$ and $b_j$, respectively,
are $h(a_j)=\tilde h(a _j)+(A_i+{\rm i}B_i)\oint_{a_j}\alpha^i_a+(C_i+
{\rm i}D_i)\oint_{a_j}\beta^i_a=\tilde h(a_j)+(A_i+{\rm i}B_i)M^i{}_j+
(C_i+{\rm i}D_i)M^{g+i} {}_j$ and $h(b_j)=\tilde h(b_j)+(A_i+{\rm i}B_i)
\oint_{b_j}\alpha^i_a+ (C_i+{\rm i}D_i)\oint_{b_j}\beta^i_a=\tilde h(b
_j)+(A_i+{\rm i}B_i)M^i{}_{g+j}+(C_i+{\rm i}D_i)M^{g+i}{}_{g+j}$. Since 
$M^I{}_J$, $I,J=1,\ldots , 2g$, is nonsingular, there exist (uniquely 
determined) constants $A_i$, $B_i$, $C_i$ and $D_i$ such that $h(a_i)=0$ 
and $h(b_i)=0$.] Then for some globally defined real 1-forms $V_a$, $Z_a$
on $\S$ one has $\edth_a\phi={}_0\edth_a\phi+p(V_a+{\rm
i}Z_a)\phi+q(V_a-{\rm i}Z_a) \phi$ for any $\phi\in
E^\infty(p,q)$. Let $\phi_0\in E^\infty(p,q)$ be constant with respect
to ${}_0\edth_a$, which is necessarily nowhere vanishing. Then any
smooth cross section of $E(p,q)$ can be written as $\phi=f\phi_0$ for
some $f\in E^\infty(0,0)$, thus its edth- and edth-prime derivatives
are $\edth\phi=(\edth f+pm^a(V_a+{\rm i}Z_a)f+q m^a(V_a-{\rm
i}Z_a)f)\phi_0$ and $\edth'\phi=(\edth' f+p\bar m^a(V_a+ {\rm
i}Z_a)f+q\bar m^a(V_a-{\rm i}Z_a)f)\phi_0$, respectively. However, by
\eqref{(4.3)} there is a one parameter family of solutions to $\edth
f+pm^a (V_a+{\rm i}Z_a)f+qm^a(V_a-{\rm i}Z_a)f=0$, and another family
of solutions to $\edth' f'+p\bar m^a(V_a+{\rm i}Z_a)f'+q\bar
m^a(V_a-{\rm i}Z_a)f'=0$, therefore $\dimker\,\edth_{(p,q)}\geq1$ and
$\dimker\,\edth'_{(p,q)}\geq1$. Then, however, \eqref{(4.4)} implies
that
\begin{equation}
\dimker\,\edth_{(p,q)}=\dimker\,\edth'_{(p,q)} =1 \hskip 12pt {\rm if}
\hskip 12pt (p-q)(1-g)=0. \label{(4.5)}
\end{equation}
Therefore, in particular, on the tori the edth and edth-prime
operators have one dimensional kernels, independently of the type
$(p,q)$ of the line bundle. These kernel and cokernel spaces can be
visualized by drawing the sequence of the edth operators between the
various line bundles (Fig.~\ref{fig:torus}).

\begin{figure}[htbp]
\begin{center}
  \setlength{\unitlength}{1mm}
\begin{picture}(120,50)(-60,-10)
\thicklines
\put(0,0){\line(0,1){30}}
\put(20,0){\line(0,1){30}}
\put(40,0){\line(0,1){30}}
\put(-20,0){\line(0,1){30}}
\put(-40,0){\line(0,1){30}}
\thinlines

\put(-60,-10){\makebox(0,0)[b]{$\cdots$}}
\put(-40,-10){\makebox(0,0)[b]{$E(-2,2)$}}
\put(-20,-10){\makebox(0,0)[b]{$E(-1,1)$}}
\put(0,-10){\makebox(0,0)[b]{$E(0,0)$}}
\put(20,-10){\makebox(0,0)[b]{$E(1,-1)$}}
\put(40,-10){\makebox(0,0)[b]{$E(2,-2)$}}
\put(60,-10){\makebox(0,0)[b]{$\cdots$}}

\put(-50,0){\vector(1,0){20}}
\put(-30,0){\vector(1,0){20}}
\put(-10,0){\vector(1,0){20}}
\put( 10,0){\vector(1,0){20}}
\put( 30,0){\vector(1,0){20}}
\put( 50,0){\line(1,0){10}}

\put(-30,-4){\makebox(0,0)[b]{$\edth$}}
\put(-10,-4){\makebox(0,0)[b]{$\edth$}}
\put(10,-4){\makebox(0,0)[b]{$\edth$}}
\put(30,-4){\makebox(0,0)[b]{$\edth$}}
\put(50,-4){\makebox(0,0)[b]{$\edth$}}

\put( 50,30){\vector(-1,0){20}}
\put( 30,30){\vector(-1,0){20}}
\put( 10,30){\vector(-1,0){20}}
\put(-10,30){\vector(-1,0){20}}
\put(-30,30){\vector(-1,0){20}}
\put(-50,30){\line(-1,0){10}}

\put(-50,31){\makebox(0,0)[b]{$\edth^\dagger$}}
\put(-30,31){\makebox(0,0)[b]{$\edth^\dagger$}}
\put(-10,31){\makebox(0,0)[b]{$\edth^\dagger$}}
\put(10,31){\makebox(0,0)[b]{$\edth^\dagger$}}
\put(30,31){\makebox(0,0)[b]{$\edth^\dagger$}}

\linethickness{1pt}

\put(40,25){\circle*{1}}
\put(40,25){\line(4,-5){20}}
\put(40,25){\line(0,-1){25}}
\put(39,12){\makebox(0,0)[r]{5}}

\put(20,15){\circle*{1}}
\put(20,15){\line(4,-3){20}}
\put(20,15){\line(0,-1){15}}
\put(19, 7){\makebox(0,0)[r]{3}}

\put(0,5){\circle*{1}}
\put(0,5){\line(4,-1){20}}
\put(0,5){\line(0,-1){5}}
\put(-1,2){\makebox(0,0)[r]{1}}

\put( 0,25){\circle*{1}}
\put( 0,25){\line(-4,1){20}}
\put( 0,25){\line(0,1){5}}
\put( 1,27){\makebox(0,0)[l]{1}}

\put(-20,15){\circle*{1}}
\put(-20,15){\line(-4,3){20}}
\put(-20,15){\line(0,1){15}}
\put(-19,22){\makebox(0,0)[l]{3}}

\put(-40,5){\circle*{1}}
\put(-40,5){\line(-4,5){20}}
\put(-40,5){\line(0,1){25}}
\put(-39,15){\makebox(0,0)[l]{5}}

\end{picture}
\end{center}
\caption{\label{fig:tensorsphere}The tensorial series for g=0 and 
  vanishing boost weight: $p+q=0$. The operator $\edth$ is injective 
  only for negative spin weights ($p<q$), and surjective only for 
  non-negative spin weights ($p\geq q$). Thus in the tensorial series 
  $\edth$ and $\edth'$ are never isomorphisms.}
\end{figure}
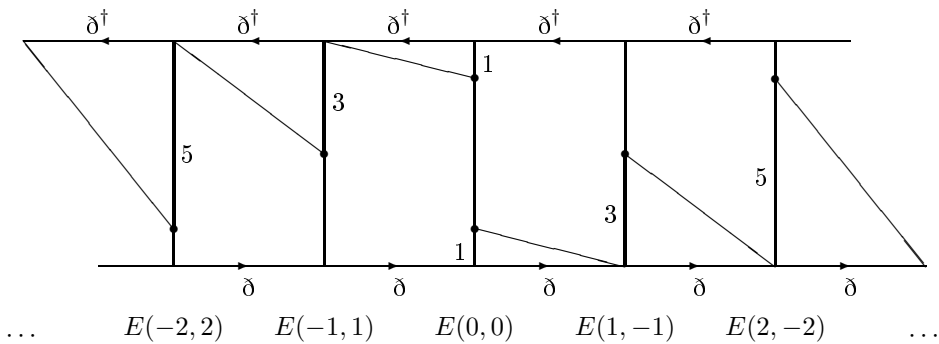

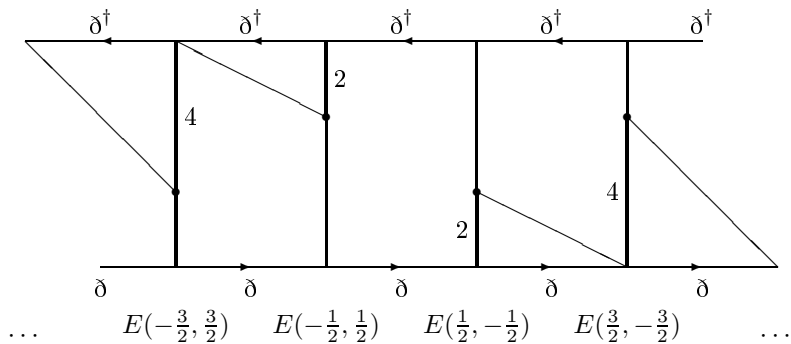
\begin{figure}[htbp]
\begin{center}
  \setlength{\unitlength}{1mm}
\begin{picture}(120,50)(-60,-10)
\thicklines
\put(10,0){\line(0,1){30}}
\put(30,0){\line(0,1){30}}
\put(-10,0){\line(0,1){30}}
\put(-30,0){\line(0,1){30}}

\put(-50,-10){\makebox(0,0)[b]{$\cdots$}}
\put(-30,-10){\makebox(0,0)[b]{$E(-\tfrac{3}{2}, \tfrac{3}{2})$}}
\put(-10,-10){\makebox(0,0)[b]{$E(-\tfrac{1}{2}, \tfrac{1}{2})$}}
\put( 10,-10){\makebox(0,0)[b]{$E( \tfrac{1}{2},-\tfrac{1}{2})$}}
\put( 30,-10){\makebox(0,0)[b]{$E( \tfrac{3}{2},-\tfrac{3}{2})$}}
\put( 50,-10){\makebox(0,0)[b]{$\cdots$}}

\thinlines

\put(-40,0){\vector(1,0){20}}
\put(-20,0){\vector(1,0){20}}
\put( 0,0){\vector(1,0){20}}
\put( 20,0){\vector(1,0){20}}
\put( 40,0){\line(1,0){10}}

\put(-40,-4){\makebox(0,0)[b]{$\edth$}}
\put(-20,-4){\makebox(0,0)[b]{$\edth$}}
\put(  0,-4){\makebox(0,0)[b]{$\edth$}}
\put( 20,-4){\makebox(0,0)[b]{$\edth$}}
\put( 40,-4){\makebox(0,0)[b]{$\edth$}}

\put( 40,30){\vector(-1,0){20}}
\put( 20,30){\vector(-1,0){20}}
\put(  0,30){\vector(-1,0){20}}
\put(-20,30){\vector(-1,0){20}}
\put(-40,30){\line(-1,0){10}}

\put(-40,31){\makebox(0,0)[b]{$\edth^\dagger$}}
\put(-20,31){\makebox(0,0)[b]{$\edth^\dagger$}}
\put(  0,31){\makebox(0,0)[b]{$\edth^\dagger$}}
\put( 20,31){\makebox(0,0)[b]{$\edth^\dagger$}}
\put( 40,31){\makebox(0,0)[b]{$\edth^\dagger$}}

\linethickness{1pt}

\put(10,10){\circle*{1}}
\put(10,10){\line(2,-1){20}}
\put(10,10){\line(0,-1){10}}
\put( 9,5){\makebox(0,0)[r]{2}}

\put(30,20){\circle*{1}}
\put(30,20){\line(1,-1){20}}
\put(30,20){\line(0,-1){20}}
\put(29,10){\makebox(0,0)[r]{4}}

\put(-10,20){\circle*{1}}
\put(-10,20){\line(-2,1){20}}
\put(-10,20){\line(0,1){10}}
\put(- 9,25){\makebox(0,0)[l]{2}}

\put(-30,10){\circle*{1}}
\put(-30,10){\line(-1,1){20}}
\put(-30,10){\line(0,1){20}}
\put(-29,20){\makebox(0,0)[l]{4}}

\end{picture}
\end{center}
\caption{\label{fig:spinorsphere}The spinorial series for $g=0$ in the 
  case of vanishing boost weight. $\edth$ is injective for negative spin 
  weights, and surjective for spin weights greater than or equal to 
  $-\tfrac{1}{2}$. Thus $\edth$ is isomorphism precisely between $E(p,
  p+1)$ and $E(p+1,p)$.}

\end{figure}

Although $\dimker\,\edth_{(p,q)}$ and $\dimker\,\edth'_{(p,q)}$ have 
been calculated for $g=0$ in the Appendix of 
\cite{szabados94:_two_sen_quasi_loc}, here we give a considerably 
simpler calculation of them. Thus let $p\in\RR$ and $n\in\NN$. Then by 
\eqref{(4.1)} one has $\dimker\,\edth_{(-p+n,-p)}=1+n+\dimker\,\edth'
_{(p+1,p-n -1)}\geq1+n$ and $\dimker\,\edth'_{(-p,-p+n)}=1+n+\dim\ker\,
\edth_{(p-n-1,p+1)}$ $\geq1+n$. Thus by \eqref{(4.4)}
\begin{equation}
\dimker\,\edth_{(p-n,p)}=\dimker\,\edth'_{(p,p-n)}=0  
\label{(4.6)}
\end{equation}
for any $p\in\RR$ and $n\in\NN$. Substituting this back to \eqref{(4.1)} 
we get 
\begin{equation}
\dimker\,\edth_{(p+n,p)}=\dimker\,\edth'
_{(p,p+n)}=1+n. \label{(4.7)}
\end{equation}
\eqref{(4.5)} (for $q=p$), \eqref{(4.6)} and \eqref{(4.7)} give the
complete list of the dimension of the kernel spaces of $\edth$ and
$\edth'$ on the spheres (see also~\cite{eastwoodtod82:_edth}). The line 
bundles and the edth operators form two sequences, the one in which 
$p-q$ is even (i.e. if the spin weight is integer, the `tensorial 
sequence'), and in which $p-q$ is odd (half-integer spin weight, the 
`spinorial sequence'). These are shown by Fig.~\ref{fig:tensorsphere}
and~\ref{fig:spinorsphere} in the case of vanishing boost weight: 
$p+q=0$.

\newpage

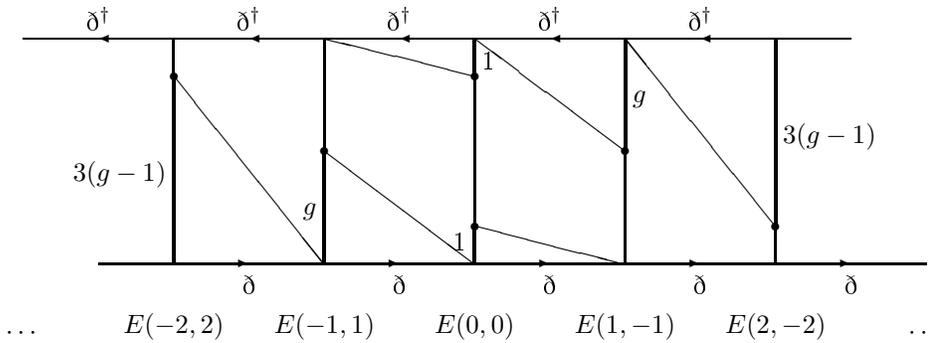
\begin{figure}[htbp]
\begin{center}
  \setlength{\unitlength}{1mm}
\begin{picture}(120,50)(-60,-10)
\thicklines
\put(0,0){\line(0,1){30}}
\put(20,0){\line(0,1){30}}
\put(40,0){\line(0,1){30}}
\put(-20,0){\line(0,1){30}}
\put(-40,0){\line(0,1){30}}
\thinlines

\put(-60,-10){\makebox(0,0)[b]{$\cdots$}}
\put(-40,-10){\makebox(0,0)[b]{$E(-2,2)$}}
\put(-20,-10){\makebox(0,0)[b]{$E(-1,1)$}}
\put(  0,-10){\makebox(0,0)[b]{$E(0,0)$}}
\put( 20,-10){\makebox(0,0)[b]{$E(1,-1)$}}
\put( 40,-10){\makebox(0,0)[b]{$E(2,-2)$}}
\put( 60,-10){\makebox(0,0)[b]{$\cdots$}}

\put(-50,0){\vector(1,0){20}}
\put(-30,0){\vector(1,0){20}}
\put(-10,0){\vector(1,0){20}}
\put( 10,0){\vector(1,0){20}}
\put( 30,0){\vector(1,0){20}}
\put( 50,0){\line(1,0){10}}

\put(-30,-4){\makebox(0,0)[b]{$\edth$}}
\put(-10,-4){\makebox(0,0)[b]{$\edth$}}
\put(10,-4){\makebox(0,0)[b]{$\edth$}}
\put(30,-4){\makebox(0,0)[b]{$\edth$}}
\put(50,-4){\makebox(0,0)[b]{$\edth$}}

\put( 50,30){\vector(-1,0){20}}
\put( 30,30){\vector(-1,0){20}}
\put( 10,30){\vector(-1,0){20}}
\put(-10,30){\vector(-1,0){20}}
\put(-30,30){\vector(-1,0){20}}
\put(-50,30){\line(-1,0){10}}

\put(-50,31){\makebox(0,0)[b]{$\edth^\dagger$}}
\put(-30,31){\makebox(0,0)[b]{$\edth^\dagger$}}
\put(-10,31){\makebox(0,0)[b]{$\edth^\dagger$}}
\put(10,31){\makebox(0,0)[b]{$\edth^\dagger$}}
\put(30,31){\makebox(0,0)[b]{$\edth^\dagger$}}

\linethickness{1pt}

\put(-40,25){\circle*{1}}
\put(-40,25){\line(4,-5){20}}
\put(-40,25){\line(0,-1){25}}
\put(-41,12){\makebox(0,0)[r]{$3(g-1)$}}

\put(-20,15){\circle*{1}}
\put(-20,15){\line(4,-3){20}}
\put(-20,15){\line(0,-1){15}}
\put(-21,7){\makebox(0,0)[r]{$g$}}

\put(0,5){\circle*{1}}
\put(0,5){\line(4,-1){20}}
\put(0,5){\line(0,-1){5}}
\put(-1,3){\makebox(0,0)[r]{1}}

\put(  0,25){\circle*{1}}
\put(  0,25){\line(-4,1){20}}
\put(  0,25){\line(0,1){5}}
\put(  1,27){\makebox(0,0)[l]{1}}

\put( 20,15){\circle*{1}}
\put( 20,15){\line(-4,3){20}}
\put( 20,15){\line(0,1){15}}
\put( 21,22){\makebox(0,0)[l]{$g$}}

\put( 40,5){\circle*{1}}
\put( 40,5){\line(-4,5){20}}
\put( 40,5){\line(0,1){25}}
\put( 41,17){\makebox(0,0)[l]{$3(g-1)$}}

\end{picture}
\end{center}
\caption{\label{fig:tensorhigher}The tensorial series for $g>1$. Note, 
  that we have restricted ourselves to the case of vanishing boost 
  weight. }

\end{figure}

\begin{figure}[htbp]
\begin{center}
  \setlength{\unitlength}{1mm}
\begin{picture}(120,50)(-60,-10)
\thicklines
\put(10,0){\line(0,1){30}}
\put(30,0){\line(0,1){30}}
\put(-10,0){\line(0,1){30}}
\put(-30,0){\line(0,1){30}}

\put(-50,-10){\makebox(0,0)[b]{$\cdots$}}
\put(-30,-10){\makebox(0,0)[b]{$E(-\tfrac{3}{2}, \tfrac{3}{2})$}}
\put(-10,-10){\makebox(0,0)[b]{$E(-\tfrac{1}{2}, \tfrac{1}{2})$}}
\put( 10,-10){\makebox(0,0)[b]{$E( \tfrac{1}{2},-\tfrac{1}{2})$}}
\put( 30,-10){\makebox(0,0)[b]{$E( \tfrac{3}{2},-\tfrac{3}{2})$}}
\put( 50,-10){\makebox(0,0)[b]{$\cdots$}}

\thinlines

\put(-40,0){\vector(1,0){20}}
\put(-20,0){\vector(1,0){20}}
\put( 0,0){\vector(1,0){20}}
\put( 20,0){\vector(1,0){20}}
\put( 40,0){\line(1,0){10}}

\put(-40,-4){\makebox(0,0)[b]{$\edth$}}
\put(-20,-4){\makebox(0,0)[b]{$\edth$}}
\put(  0,-4){\makebox(0,0)[b]{$\edth$}}
\put( 20,-4){\makebox(0,0)[b]{$\edth$}}
\put( 40,-4){\makebox(0,0)[b]{$\edth$}}

\put( 40,30){\vector(-1,0){20}}
\put( 20,30){\vector(-1,0){20}}
\put(  0,30){\vector(-1,0){20}}
\put(-20,30){\vector(-1,0){20}}
\put(-40,30){\line(-1,0){10}}

\put(-40,31){\makebox(0,0)[b]{$\edth^\dagger$}}
\put(-20,31){\makebox(0,0)[b]{$\edth^\dagger$}}
\put(  0,31){\makebox(0,0)[b]{$\edth^\dagger$}}
\put( 20,31){\makebox(0,0)[b]{$\edth^\dagger$}}
\put( 40,31){\makebox(0,0)[b]{$\edth^\dagger$}}

\linethickness{1pt}

\put(-30,20){\circle*{1}}
\put(-30,20){\line(1,-1){20}}
\put(-30,20){\line(0,-1){20}}
\put(-31,10){\makebox(0,0)[r]{$2(g-1)$}}

\put(-10,10){\circle*{1}}
\put(-10,10){\line(2,-1){20}}
\put(-10,10){\line(0,-1){10}}
\put(-11,5){\makebox(0,0)[r]{$g-1$}}

\put( 10,20){\circle*{1}}
\put( 10,20){\line(-2,1){20}}
\put( 10,20){\line(0,1){10}}
\put( 11,25){\makebox(0,0)[l]{$g-1$}}

\put( 30,10){\circle*{1}}
\put( 30,10){\line(-1,1){20}}
\put( 30,10){\line(0,1){20}}
\put( 31,20){\makebox(0,0)[l]{$2(g-1)$}}

\end{picture}
\end{center}
\caption{\label{fig:spinorhigher}The spinorial series for $g>1$ in the 
  case of vanishing boost weight. }

\end{figure}
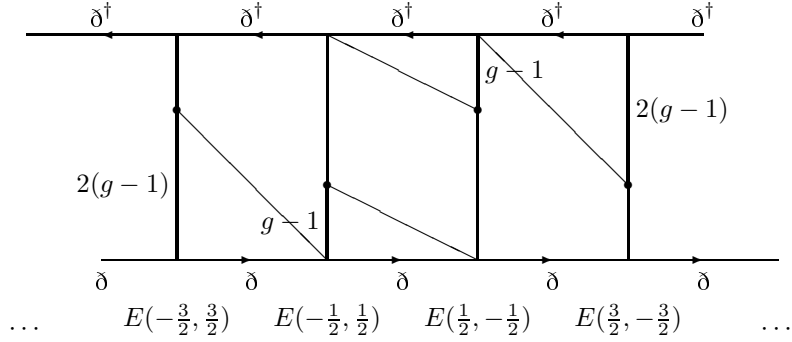

Finally, let $g\geq2$. By \eqref{(4.1)} 
\begin{equation}
\begin{split}
  \dimker\,\edth_{(p,p+n+1)}&=n(g-1)+\dimker\, \edth'_{(-p-n,-p-1)},\\
  \dimker\,\edth'_{(p,p-n-1)}&=n(g-1)+\dimker\,
  \edth_{(-p+n,-p+1)}
\end{split}
\label{(4.8)}
\end{equation}
for any $n\in\{0\}\cup\NN$ and $p\in\RR$. Then for $n\geq2$
and any $p\in\RR$ these, and for $n=1$ and any $p\in\RR$ these
and \eqref{(4.5)} imply $\dimker\,\edth_{(p,p+n+1)}\geq2$ and
$\dimker\,\edth'_{(p,p-n-1)}\geq2$. Therefore, by \eqref{(4.4)} it
follows that
$\dimker\,\edth_{(p,p-n-1)}=\dimker\,\edth'_{(p,p+n+1)}=0$ for any
$n\in\NN$ and $p\in\RR$. However, this implies that
\begin{equation}
\dimker\,\edth_{(p,p-n)}=\dimker\,\edth'_{(p,p+n)}=0 \label{(4.9)}
\end{equation}
for any $n\in\NN$ and $p\in\RR$, too. To see this,
suppose, on the contrary, that e.g. $\dimker\,\edth_{(p,p-1)}\geq1$
for some $p\in\RR$. Then by \eqref{(4.2)} this would imply that
$\dimker\,\edth_{(2p,2p-2)}\geq1$. Substituting \eqref{(4.9)} back to
\eqref{(4.8)} we obtain
\begin{equation}
\dimker\,\edth_{(p,p+n+1)}=\dimker\,\edth'_{(p,p-n-1)}=\left\{
  \begin{array}{rl}
    g & \mbox{if } n=1\\ 
    n(g-1) & \mbox{if } n \ge 2
  \end{array}\right.
\label{(4.10)}
\end{equation}
for any $n\in\NN$ and $p\in\RR$. Note that for $g\geq2$ the 
Riemann--Roch theorem, more precisely, \eqref{(4.8)} for $n=0$, states 
only that two unknown dimensions are equal: $\dimker\,\edth_{(p,p+1)}=
\dimker\,\edth'_{(-p,-p-1)}$. Thus to calculate them let us choose a 
real harmonic 1-form $\omega_a$ and define the map $\bar\omega:\ker\,
\edth_{(0,1)}\rightarrow\ker\,\edth_{(-1,2)}:\phi\mapsto\bar\omega\phi$. 
Obviously, this is injective. By \eqref{(4.10)} $\dimker\,\edth_{(-1,2)}
=2(g-1)$, and ${\rm deg}(\nu)=c_1(-1,2)= 3(g-1)$ for any $\nu\in\ker\,
\edth_{(-1,2)}$, whilst ${\rm deg}(\bar \omega)=c_1(-1,1)=2(g-1)$. Thus 
the quotient $\nu/\bar\omega$ can be holomorphic only if the zeros of 
$\bar\omega$ are compensated by the zeros of $\nu$, hence the map $\bar
\omega$ is not surjective. Its cokernel is $2(g-1)$ dimensional, and 
hence $\dimker\,\edth_{(0,1)}=\dimker\,\edth'_{(1,0)}=g-1$. Finally, 
since by $\dimker\,\edth_{(p,p)}=1$ we have $\dimker\,\edth_{(p,p+1)}
\geq {\rm max}\{\dimker\,\edth_{(0,1)},\,\dimker\, \edth_{(p,p)}\}=g-1$. 
On the other hand, by $\dimker\,\edth_{(-p,-p)}=1$ we have $g-1=\dimker\,
\edth_{(0,1)}\geq{\rm max}\,\{\dimker\,\edth_{(p,p+1)},\,\dimker\,\edth
_{(-p,-p)}\}=\dimker\,\edth_{(p,p+1)}$. Therefore,
\begin{equation}
\dimker\,\edth_{(p,p+1)}=\dimker\,\edth'
_{(p+1,p)}=g-1  \label{(4.11)}
\end{equation}
for any $p\in\RR$. For $g\geq2$ \eqref{(4.5)}, 
\eqref{(4.9)}--\eqref{(4.11)} is the complete list of the dimension of 
the kernel spaces of the edth and edth-prime operators. The tensor- and 
spinor sequences of the edth operators are shown by 
Fig.~\ref{fig:tensorhigher} and~\ref{fig:spinorhigher}, respectively.  

\section{Discussion}

The figures show clearly that $g=1$ is a natural division between  the
spherical case ($g=0$) and the higher genus surfaces. While on the
sphere, the edth operator has non-trivial kernels for $p\ge q$, on the
surfaces with higher genus the kernels are non-trivial for $p\le
q$. This is obviously related to the fact that on the sphere there
exist conformal Killing vectors but no harmonic forms, while for $g\ge
2$ there are harmonic forms but no conformal Killing vectors. On the
torus there is exactly one complex one of each. In fact, by
\eqref{(4.5)}, \eqref{(4.7)} and \eqref{(4.9)} the number of the
independent real conformal Killing vectors is six on $S^2$, two on
$S^1\times S^1$ and zero on surfaces with genus $g\geq2$. Similarly,
by \eqref{(4.5)}, \eqref{(4.7)} and \eqref{(4.10)}
$\dimker\,\edth'_{(1,-1)}=g$, i.e. there are $g$ holomorphic 1-forms
on a surface with genus $g$ which can be combined with their
(anti-holomorphic) complex conjugates to yield the $2g$ real harmonic
1-forms whose existence is guaranteed by the Hodge decomposition. 

It is well known that in the case $g=0$ the kernel of
$\edth_{(\2,-\2)}$ serves as a building block for the kernels of
$\edth_{(s,-s)}$ with $2s \in \NN$ in the sense that if
$\{\alpha_1,\alpha_2\}$ is a basis in $\ker\,\edth_{(\2,-\2)}$ then
$\{\alpha^{2s}_1,\alpha^{2s-1}_1\alpha_2, \ldots,\alpha^{2s}_2\}$ is a
basis in $\ker\,\edth_{(s,-s)}$. Similarly, on the torus $\dimker\,
\edth_{(p,q)}=1$ for any $p,q$, thus if $\alpha\in \ker\edth_{(\2,-\2)}$
and $\beta\in \ker\edth_{(-\2,\2)}$, then $\alpha^{2s}$ spans
$\ker\,\edth_{(s, -s)}$ and $\beta^{2s}$ spans
$\ker\,\edth_{(-s,s)}$. If, however, $g=2$ then
$\dimker\,\edth_{(-\2,\2)}=1$, but $\dimker\,\edth_{(-s,s)}\geq2$ for
$s\geq1$, $2s\in\NN$, and hence the elements of $\ker\,\edth_{(-s,s)}$
cannot be generated by the single independent element of
$\ker\,\edth_{(-\2,\2)}$.

To understand the geometric roots of this difference between the
$g\leq1$ and $g\geq2$ cases, recall that the elements of
$\ker\,\edth_{(1,-1)}$ correspond to globally defined conformal
Killing vectors, which generate global group actions on $\S$. Then
$\ker\,\edth_{(\2,-\2)}$ is the representation space of the double
covering group of this symmetry group, and hence any irreducible
representation of the symmetry group is build from
$\edth_{(\2,-\2)}$. On $S^2$ this group is $SL(2,\CC)$, on $S^1\times
S^1$ it is $U(1)\times U(1)$, but there is no such group on
higher-genus 2-surfaces. The harmonic 1-forms, which correspond to the
elements of $\ker\,\edth_{(-1,1)}$, do not generate any such group
action on $\S$, and in lack of such a group structure the spaces
$\ker\,\edth_{(-\2,\2)}$ and $\ker\,\edth_{(-s,s)}$ cannot be expected
to be related as (different weight) representation spaces of a group.

Yet, it is obvious that for any holomorphic 1-form $\omega \in
\ker\edth_{(-1,1)}$ and any $\alpha\in\ker \edth_{(-s,s)}$ we have
$\omega\alpha \in\ker \edth_{(-s-1,s+1)}$ so that the holomorphic
1-forms do in fact map the kernels into each other. So the question
arises as to whether one can obtain all the elements in $\ker
\edth_{(-s,s)}$ for $s\in \NN$ as linear combinations of the $s$-fold
products of the $g$ holomorphic 1-forms. However, this cannot be true
in general because it is easy to see that in the case of
hyper-elliptic Riemann surfaces with genus $g\ge3$
(see~\cite{farkaskra93:_rieman_surfac}) these products are not
sufficient to span the entire kernel because they satisfy too many
linear relations. But these are the only exceptions: a rather deep
result in the theory of Riemann surfaces of higher genus, the theorem
of Noether~\cite{farkaskra93:_rieman_surfac} states that, except for
these special cases, the kernels $\ker \edth_{(-s,s)}$ in the
tensorial series, i.e., for $s\in \NN$ are generated by the $s$-fold
products of holomorphic 1-forms. It would be interesting to have a
similar result for the spinorial series.

To solve the equation $\edth_{(p,q)}\phi=\omega$ for $\phi$ with given 
$\omega$, the cross section $\omega\in E^\infty(p+1,q-1)$ must belong 
to the range of $\edth_{(p,q)}$, i.e. $\omega$ must be orthogonal (with 
respect to \eqref{(3.1)}) to $\ker\,(\edth_{(p,q)})^\dagger$. Then there 
is a unique solution $\phi$ if that is chosen to be orthogonal to $\ker\,
\edth_{(p,q)}$. In fact, $\edth$ is a continuous linear operator with 
respect to the standard Sobolev norms on $E^\infty(p,q)$ and $E^\infty
(p+1,q-1)$, and hence it is a Fredholm operator. Then this criterion of 
the solvability of $\edth_{(p,q)}\phi=\omega$ is just the Fredholm 
alternative theorem (see~\cite{besse87:_einst_manif}). 

In particular, 
to answer the question posed in the introduction on the vanishing of
the Weyl tensor on the conformal boundaries with higher genus
topology, we note that this leads to the equation $\edth\phi=0$ for $\phi
\in E^\infty(4,0)$. While for $g=0$ the appropriate kernel is trivial, 
this is not the case if $g\ge1$. Then there are non-trivial solutions so
that the Weyl tensor does not necessarily vanish on toroidal (and other
higher genus) null-infinities.

\section*{Acknowledgments}

JF wishes to thank the Theoretical Department of the Research Institute 
for Particle and Nuclear Physics in Budapest for support, while LBSz 
would like to thank the MPI for Mathematics in the Sciences for its 
hospitality. This work was partially supported by the Hungarian 
Scientific Research Fund grant OTKA T030374. 


\end{document}